\documentclass[%
 reprint,
 amsmath,amssymb,
 aps,
 showkeys,
]{revtex4-2}

\usepackage{graphicx}
\usepackage{dcolumn}
\usepackage{bm}
\usepackage[hidelinks]{hyperref}

\usepackage{listings}
\begin{document}

\preprint{[arXiv ID]}

\title{Daylight-Referenced Efficiency (DRE):\\A Radiometric Measure of Electric Light Source Efficiency}
\author{Peter Veto}
\affiliation{Budapest, Hungary}
\date{26 October 2026}

\begin{abstract}
Photopic luminous efficacy ($\text{lm/W}$) weights optical power strictly against human eye sensitivity $V(\lambda)$, creating a systemic metrological penalty for electric light sources engineered with broad spectral output in the ultraviolet ($280\text{--}380\text{ nm}$) or infrared ($780\text{--}4000\text{ nm}$) bands. Similarly, metrics evaluating chromaticity, color quality, or melanopic pathways assess photoreceptor responses strictly within the visual band, offering no measure of broad-spectrum physical radiant power conversion. To resolve this limitation, we introduce Daylight-Referenced Efficiency ($\eta_{\text{DRE}}$), a radiometric framework that evaluates how efficiently an electric light source converts electrical power into an optical spectrum matching natural terrestrial daylight across domain $\Omega = [280\text{ nm}, 4000\text{ nm}]$. The framework factorizes performance into primary physical quantities: domain-restricted Wall-Plug Efficiency ($\eta_{\text{wall}}$ [$\text{W}_{\text{optical}}/\text{W}_{\text{elec}}$]), measuring unweighted electrical-to-optical conversion, and Daylight Spectral Fidelity ($\eta_{\text{spec}}$ [unitless, $0\text{--}1$]), quantifying geometric shape match to the ASTM G173 AM1.5G solar reference via total variation distance ($1 - \text{TVD}$). Multiplying total optical power by spectral fidelity yields the absolute Daylight-Referenced Output Power ($P_{\text{dr}}$ [$\text{W}$]), establishing a direct physical flux quantity for lighting specification. Dividing by electrical drive power gives the summary metric Daylight-Referenced Efficiency ($\eta_{\text{DRE}} = \eta_{\text{wall}} \times \eta_{\text{spec}}$ [$\text{W}/\text{W}_{\text{elec}}$]), measuring net reference-conforming radiant output per electrical Watt consumed. The mathematical architecture is reference-agnostic, domain-extensible, and maintains provable lower-bound monotonicity for $\eta_{\text{wall}}$ and $\eta_{\text{DRE}}$ under spectral truncation. Open-source reference software is provided to facilitate deployment and standardized rating of broad-spectrum light sources.
\end{abstract}


\maketitle


\section{Introduction}

The standard spectral luminous efficiency function for photopic human vision --- denoted as $V(\lambda)$ --- has been the foundation of visual photometry for over a century \cite{CIE1924}. The $V(\lambda)$ sensitivity curve peaks at $555\text{ nm}$ and truncates outside the visual domain ($360\text{ nm}\text{--}830\text{ nm}$). 

In optical engineering, $V(\lambda)$ forms the baseline for quantifying the electrical energy efficiency of light sources via luminous efficacy, expressed in lumens per electrical Watt ($\text{lm/W}$). Luminous efficacy measures the total emitted luminous flux relative to total active input electrical power. By weighting a source's spectral power distribution against $V(\lambda)$, $\text{lm/W}$ evaluates approximate visual brightness per unit of electrical power for a standard light-adapted human observer.

\subsection{Limitations of Visual-Centric Metrics}
While photopic luminous efficacy provides a benchmark for visual task performance, it is structurally constrained to human visual cone photoreception. Similarly, metrics evaluating chromaticity \cite{CIE133_1995, ANSI_TM30_20}, melanopsin-mediated responses \cite{CIES026, Rea2012}, or even similarity to a daylight reference \cite{IEC60904_9_2020, SMPTE_ST2122_2020, Bridgelux_ASD_2019, Acosta2018DSI} are limited by one or more of the following restraints: they focus on retinal photoreceptor responses, they remain within the visual band, or they do not extend beyond a truncated portion of the near-infrared range. Furthermore, none of these constructs quantify physical radiant power conversion across extended optical wavebands, evaluated against natural terrestrial daylight.

\subsection{Broad-Spectrum Emitters and Metrological Constraints}
Electric lighting technologies increasingly feature spectral power distributions extending across and beyond the visual spectrum, targeting spectral matching with natural daylight for architectural, photovoltaic, horticultural, and physiological applications \cite{waveform_fullspectrum_br30_2026, seoul_sunlike_2026, telelumen_spectral_2026, yuji_sunwaveplus_2026, phosphortech_nir_2026, dimlux_xplore_2026, restful_bedtimebulb_v2_2026, spectrasol_technology_2026, nira_lighting_a19_2026}. These light sources incorporate photon output in the ultraviolet ($280\text{ nm}\text{--}380\text{ nm}$) or infrared ($780\text{ nm}\text{--}4000\text{ nm}$) domains, sometimes motivated by visual \cite{barrett2026led, bergquist2024pitfalls, kwak2024vision, kotulak1994does, swiatczak2024inhibiting, drew2012chromatic, swiatczak2022myopia, xiong2017time} or non-visual physiological mechanisms \cite{schierz2019light, sanderson2018inhibitory, galache2024photobiomodulation, hou2024light, bao2025green, cidral2025violet, noda2025violet, torii2022short, yadav2017noninvasive, roddick2024effects}.

Evaluating such broad-spectrum emitters using conventional photopic efficacy introduces a systemic metrological penalty. Because $V(\lambda)$ assigns zero weight to radiation outside the visual band, electrical energy converted into ultraviolet or infrared flux is treated as energy loss. Similarly, wavelengths toward both ends of the visible range are weighted less than those near its center. Consequently, sources engineered for broad spectral continuity exhibit suppressed $\text{lm/W}$ values despite high total radiant efficiency \cite{Schubert2006}. As commercial luminaires are increasingly engineered and marketed on claims of daylight-like spectral output, objective metrology requires evaluating their emission directly against an explicit terrestrial solar reference. Furthermore, evaluating energy trade-offs between architectural daylighting and electric lighting systems requires a common radiometric measure capable of accounting for the full spectral extent of solar radiation. Currently, no standardized metric quantifies how efficiently an electric light source converts electrical power into an optical spectrum that matches physical terrestrial daylight.

\subsection{Proposed Solution: Daylight-Referenced Efficiency (\texorpdfstring{$\eta_{\text{DRE}}$}{DRE})}
To address this metrological gap, we propose Daylight-Referenced Efficiency (\texorpdfstring{$\eta_{\text{DRE}}$}{DRE}), expressed as Daylight-Referenced Output Power per electrical input power ($\text{P}_{\text{dr}}/\text{P}_{\text{elec}}$). The framework factorizes performance into two independent physical terms, as well as a single combined summary efficiency metric:
\begin{enumerate}
    \item \textbf{Daylight Spectral Fidelity (\texorpdfstring{$\eta_{\text{spec}}$}{eta\_spec}):} A unitless ratio ($0\text{--}1$) evaluating the spectral shape match between the emitted light profile and a terrestrial solar reference.
    \item \textbf{Wall-Plug Efficiency (\texorpdfstring{$\eta_{\text{wall}}$}{eta\_wall}):} The ratio of radiant optical output power ($\text{P}_{\text{optical}}$) to active electrical input power ($\text{P}_{\text{elec}}$) restricted to evaluation domain $\Omega = [280\text{ nm}, 4000\text{ nm}]$, adopting the standard definition of radiant wall-plug efficiency \cite{Schubert2006} bounded to the solar reference spectrum.
\end{enumerate}
The product of these components ($\eta_{\text{DRE}} = \eta_{\text{wall}} \times \eta_{\text{spec}}$) yields the net rate of daylight-referenced optical output generated per unit of electrical power consumed.

\subsection{Selection and Flexibility of the Terrestrial Solar Reference}
Formulating a daylight-referenced metric requires defining an unambiguous terrestrial solar baseline. Daylight varies by location, time, and atmospheric conditions. However, using moment-by-moment measured local spectra is unfeasible for standardized hardware rating and cross-manufacturer comparisons.

Alternatively, a theoretical reference could be constructed from ensemble-averaged, population-weighted global daylight models that incorporate atmospheric data. While promising for future research, no standardized model currently provides this global spectrum; nevertheless, its shape can be expected to closely approximate existing terrestrial solar reference standards.

To ensure reproducibility and alignment with international standards, $\eta_{\text{DRE}}$ adopts the Air Mass 1.5 Global Tilt spectrum (AM1.5G, ASTM G173 / IEC 60904-3) across its entire spectral range ($280\text{ nm}\text{--}4000\text{ nm}$) as its primary reference standard \cite{ASTMG173}. AM1.5G is broadly established in solar metrology and models average mid-latitude terrestrial solar irradiance ($1000 \, \text{W/m}^2$). The evaluation domain $\Omega = [280\text{ nm}, 4000\text{ nm}]$ is selected specifically because this standard represents the most comprehensive, standardized terrestrial daylight reference dataset currently available. Crucially, the mathematical algorithms presented herein are reference-agnostic and domain-extensible: should broader daylight reference standards or expanded spectroradiometric capabilities emerge, the domain bounds of $\Omega$ can be extended seamlessly while the core protocol of DRE remains unchanged.

At the same time, characterization across the full $\Omega = [280\text{ nm}, 4000\text{ nm}]$ span presents a demanding metrological requirement for electric light sources. Spectroradiometric measurements spanning from the ultraviolet through the mid-infrared are instrumentally intensive, and few commercial lighting databases currently provide spectral power distributions extending across this complete domain.

For most broad-spectrum light sources, however, this metrological limitation does not affect the practical applicability of DRE. Where the spectral output lies within the conventional measurement range, daylight-referenced efficiencies can be compared without compromise. Thermal sources constitute a notable exception, as their spectral output may extend to 4000 nm or beyond. In such cases, measuring the shorter-wavelength portion of the spectrum and modeling the remaining long-wavelength contribution may offer a practical approach to estimating DRE across the full domain. Semiconductor sources operating at these wavelengths also exist, but their typically narrower emission spectra present a different characterization challenge; as of this writing, such sources are not used in conventional general-lighting applications.


\section{Mathematical Formulation}
\label{sec:math}

This section defines the mathematical architecture of the Daylight-Referenced Efficiency framework. The formulation separates hardware electrical-to-optical conversion from spectral profile conformance before combining them into a unified summary metric.

\subsection{Evaluation Domain and Physical Inputs}
The evaluation domain $\Omega$ corresponds to the full wavelength coverage of the AM1.5G terrestrial solar standard (ASTM G173 / IEC 60904-3):
\begin{equation}
\Omega = [\lambda_{\text{min}}, \, \lambda_{\text{max}}] = [280\text{ nm}, \, 4000\text{ nm}]
\end{equation}

While $\Omega$ is defined here by the spectral bounds of the AM1.5G standard, the mathematical architecture is domain-invariant and can accommodate extended wavelength boundaries whenever broader reference spectra or measurement datasets are employed.

The evaluation requires three primary input parameters:
\begin{enumerate}
    \item \textbf{Active Electrical Input Power ($P_{\text{elec}}$):} The active real electrical power consumed by the luminaire system during spectroradiometric measurement, expressed in electrical Watts ($\text{W}_{\text{elec}}$).
    \item \textbf{Absolute Spectral Power Distribution ($P_{\text{test}}(\lambda)$):} The absolute emitted radiant flux density of the test light source as a continuous function of wavelength $\lambda$ across $\Omega$, expressed in Watts per nanometer ($\text{W/nm}$).
    \item \textbf{Terrestrial Solar Reference Irradiance ($S_{\text{AM1.5G}}(\lambda)$):} The global tilt terrestrial solar spectral irradiance profile across $\Omega$, expressed in Watts per square meter per nanometer ($\text{W}\cdot\text{m}^{-2}\cdot\text{nm}^{-1}$).
\end{enumerate}

For $p(\lambda)$ and $s(\lambda)$ to admit valid interpretation as probability density functions in the total variation distance (TVD) formulation of Section~\ref{sec:spec_fidelity}, both spectral inputs are assumed non-negative across the evaluation domain:
\begin{equation}
P_{\text{test}}(\lambda) \geq 0 \quad \text{and} \quad S_{\text{AM1.5G}}(\lambda) \geq 0, \quad \forall \lambda \in \Omega
\end{equation}
Section~\ref{sec:protocol} discusses handling of measurement artifacts (e.g. instrument noise-floor excursions) that may violate this condition in practice.

Integration of $P_{\text{test}}(\lambda)$ across domain $\Omega$ yields the total unweighted radiant optical flux emitted by the source, $P_{\text{optical}}$:
\begin{equation}
P_{\text{optical}} = \int_{\lambda_{\text{min}}}^{\lambda_{\text{max}}} P_{\text{test}}(\lambda) \, d\lambda \quad [\text{W}_{\text{optical}}]
\end{equation}

\subsection{Normalization to Spectral Probability Densities}
To evaluate spectral shape fitting independently of absolute power magnitude, both the test source emission profile $P_{\text{test}}(\lambda)$ and reference solar profile $S_{\text{AM1.5G}}(\lambda)$ are scaled into probability density functions, $p(\lambda)$ and $s(\lambda)$. Although $P_{\text{test}}(\lambda)$ carries units of radiant flux ($\text{W/nm}$) while $S_{\text{AM1.5G}}(\lambda)$ carries units of irradiance ($\text{W}\cdot\text{m}^{-2}\cdot\text{nm}^{-1}$), self-normalization of each function by its own spectral integral cancels these respective units, yielding two dimensionally identical densities ($\text{nm}^{-1}$) that are directly comparable independent of the absolute magnitude or physical unit convention of the original measurement:

\begin{equation}
p(\lambda) = \frac{P_{\text{test}}(\lambda)}{P_{\text{optical}}} = \frac{P_{\text{test}}(\lambda)}{\int_{\lambda_{\text{min}}}^{\lambda_{\text{max}}} P_{\text{test}}(\lambda) \, d\lambda} \quad \left[\text{nm}^{-1}\right]
\end{equation}

\begin{equation}
s(\lambda) = \frac{S_{\text{AM1.5G}}(\lambda)}{\int_{\lambda_{\text{min}}}^{\lambda_{\text{max}}} S_{\text{AM1.5G}}(\lambda) \, d\lambda} \quad \left[\text{nm}^{-1}\right]
\end{equation}

By construction, both normalized density functions integrate to unity across domain $\Omega$:
\begin{equation}
\int_{\lambda_{\text{min}}}^{\lambda_{\text{max}}} p(\lambda) \, d\lambda = 1.0 \quad \text{and} \quad \int_{\lambda_{\text{min}}}^{\lambda_{\text{max}}} s(\lambda) \, d\lambda = 1.0
\end{equation}

\subsection{Daylight Spectral Fidelity (\texorpdfstring{$\eta_{\text{spec}}$}{eta\_spec})}
\label{sec:spec_fidelity}
Spectral profile mismatch between the normalized test density $p(\lambda)$ and reference solar density $s(\lambda)$ is defined via TVD:
\begin{equation}
\text{TVD}(p, s) = \frac{1}{2} \int_{\lambda_{\text{min}}}^{\lambda_{\text{max}}} |p(\lambda) - s(\lambda)| \, d\lambda
\end{equation}

The factor of $\frac{1}{2}$ normalizes the distance metric such that $0 \le \text{TVD}(p, s) \le 1$. Daylight Spectral Fidelity ($\eta_{\text{spec}}$) is defined as the complement of TVD:
\begin{equation}
\eta_{\text{spec}} = 1 - \text{TVD}(p, s) = 1 - \frac{1}{2} \int_{\lambda_{\text{min}}}^{\lambda_{\text{max}}} |p(\lambda) - s(\lambda)| \, d\lambda
\end{equation}

The parameter $\eta_{\text{spec}}$ is a unitless ratio satisfying $0 \leq \eta_{\text{spec}} \leq 1$, where $0$ corresponds to zero spectral overlap and $1$ to an identical spectral distribution to the solar reference baseline. (See also Appendix~\ref{app:minenv} for analytical proof of equivalence to minimum-envelope spectral overlap).

\subsection{Efficiency Decomposition and Summary Metric (\texorpdfstring{$\eta_{\text{DRE}}$}{eta\_DRE})}
Wall-Plug Efficiency ($\eta_{\text{wall}}$) quantifies the electrical-to-optical conversion efficiency restricted to evaluation domain $\Omega = [280\text{ nm}, 4000\text{ nm}]$, adopting the standard optoelectronic definition of radiant wall-plug efficiency \cite{Schubert2006} bounded to the terrestrial solar reference domain:
\begin{equation}
\eta_{\text{wall}} = \frac{P_{\text{optical}}}{P_{\text{elec}}} = \frac{\int_{\lambda_{\text{min}}}^{\lambda_{\text{max}}} P_{\text{test}}(\lambda) \, d\lambda}{P_{\text{elec}}} \quad \left[\frac{\text{W}_{\text{optical}}}{\text{W}_{\text{elec}}}\right]
\end{equation}

Multiplying total unweighted radiant optical power $P_{\text{optical}}$ by spectral fidelity $\eta_{\text{spec}}$ yields the Daylight-Referenced Output Power ($P_{\text{dr}}$):
\begin{equation}
\begin{split}
P_{\text{dr}} &= P_{\text{optical}} \times \eta_{\text{spec}} \\ &= P_{\text{optical}} \times \left( 1 - \text{TVD}(p, s) \right) \quad [\text{W}]
\end{split}
\end{equation}

The primary metric --- Daylight-Referenced Efficiency ($\eta_{\text{DRE}}$) --- is defined as the product of wall-plug efficiency and daylight spectral fidelity:
\begin{equation}
\begin{split}
\eta_{\text{DRE}} &= \eta_{\text{wall}} \times \eta_{\text{spec}} \\ &= \left( \frac{P_{\text{optical}}}{P_{\text{elec}}} \right) \times \left( 1 - \text{TVD}(p, s) \right) \\ &= \frac{P_{\text{dr}}}{P_{\text{elec}}} \quad \left[\frac{\text{W}}{\text{W}_{\text{elec}}}\right]
\end{split}
\end{equation}

The resulting quantity $\eta_{\text{DRE}}$ expresses the net rate of daylight-referenced radiant optical power generated per unit of electrical power consumed.


\section{Evaluation Protocol}
\label{sec:protocol}

This section summarizes the numerical procedure used to evaluate $\eta_{\text{DRE}}$ from measured or modeled spectra. Derivations, implementation details, and verification are given in Appendices~\ref{app:minenv}--\ref{app:results}.

\subsection{Reference Data and Evaluation Domain}
The solar reference $S_{\text{AM1.5G}}(\lambda)$ is the global-tilt column of the tabulated ASTM~G173-03 spectrum \cite{ASTMG173}, which is published on a non-uniform wavelength grid and resampled as described below. Photopic quantities used for comparison with luminous efficacy are computed from the tabulated CIE~1924 function $V(\lambda)$ \cite{CIE1924, CIE018_2019} (360--830~nm, taken as zero outside).  The conventional value $K_m = 683\ \text{lm/W}$ is used for maximum photopic luminous efficacy. The evaluation domain $\Omega = [280\text{ nm}, 4000\text{ nm}]$ is fixed; emission outside $\Omega$ does not contribute to $\eta_{\text{DRE}}$ by definition. Data provenance is documented in Appendix~\ref{app:numerics}.

\subsection{Numerical Procedure}
All spectra are resampled onto a common uniform grid of 1~nm spacing spanning $\Omega$ inclusively (3721 points) by monotone piecewise-cubic Hermite (PCHIP) interpolation. Unlike a cubic spline, PCHIP cannot overshoot the neighboring data values, so non-negative data yield a non-negative interpolant. No extrapolation is performed: grid points outside a spectrum's measured range are set to zero. All integrals ($P_{\text{optical}}$, the normalizations of $p$ and $s$, the TVD integral, and luminous flux) are evaluated with the composite Simpson rule on this grid. Negative values arising from measurement noise are set to zero and reported. For two of the sources examined in Appendix~\ref{app:numerics} (a modeled blackbody and a measured LED), refining the grid from 1~nm to 0.25~nm changes $\eta_{\text{DRE}}$ by less than $0.1\,\%$, and replacing Simpson's rule by the trapezoidal rule changes it by a comparable amount.

\subsection{Incomplete and Relative Spectra}
\emph{Truncated measurements.} If the measured range does not cover $\Omega$, the unmeasured region is treated as zero emission. Under any such truncation, $\eta_{\text{wall}}$ and $\eta_{\text{DRE}}$ are rigorous lower bounds of their true $\Omega$-restricted values, whereas $\eta_{\text{spec}}$ may be biased in either direction (Appendix~\ref{app:truncation}). A lower bound is informative only if little emission lies outside the measured range. Where significant emission can be expected there, as for thermal sources measured only in the visible and near-infrared, the bound may lie far below the true value and $\eta_{\text{spec}}$ describes only the measured part of the spectrum, so $\eta_{\text{DRE}}$ should not be used as a figure of merit unless the measurement is extended to cover that emission. The implementation issues a warning when a spectrum has not decayed at a range edge lying strictly inside $\Omega$, but this heuristic cannot detect emission that continues beyond an edge where the measured signal falls off for instrumental reasons (for example, detector sensitivity). The measured wavelength range should therefore always be reported alongside the result, and the plausibility of negligible emission outside it should be assessed from the source technology.

\emph{Relative spectra.} The metric requires an absolute spectral power distribution. When only a relative spectrum is available, its scale is fixed from a reported luminous efficacy and electrical input power (Appendix~\ref{app:anchoring}). The result is then conditional on those reported values, and the lower-bound statement above holds only if the truncated region lies outside the support of $V(\lambda)$.

\subsection{Verification}
The implementation is verified against closed-form limits ($\eta_{\text{spec}}=1$ for a source proportional to the reference, $\eta_{\text{spec}}=0$ for disjoint support, invariance of $\eta_{\text{spec}}$ under power scaling), against an independent quadrature of the Planck function, against the defining property of $K_m$, and against the min-envelope identity of Appendix~\ref{app:minenv}. A round trip through the full anchoring-and-resampling pipeline reproduces the published luminous efficacy of a measured LED to $0.002\,\%$; this tests the pipeline's numerical consistency, not the absolute calibration of the source (Appendix~\ref{app:verification}).

\subsection{Code Availability}
A reference implementation in Python is available at \href{https://doi.org/10.5281/zenodo.23669256}{10.5281/zenodo.23669256}.


\section{Application Examples}
\label{sec:examples}

This section applies the protocol of Section~\ref{sec:protocol} to measured spectra of commercial lamps and to one modeled source. Its purpose is to illustrate the quantities defined in Section~\ref{sec:math}, not to rank products.

\subsection{Sources}
\label{sec:sources}
\emph{Measured lamps.} Spectra of commercial lamps were taken from the Lamp Spectral Power Distribution Database (LSPDD) \cite{LSPDD}, which provides for each lamp a relative spectral power distribution together with the reported electrical power and photopic luminous efficacy. Sixteen records with both quantities reported were evaluated: ten LED lamps, two compact fluorescent lamps (CFL), one incandescent lamp, two halogen lamps, and one metal-halide (MH) high-bay luminaire, with calculated correlated color temperatures of 2400--5155~K and electrical powers of 4--400~W. Each spectrum was anchored to an absolute scale as described in Appendix~\ref{app:anchoring}. Four lamps are discussed in detail (Table~\ref{tab:featured}, Fig.~\ref{fig:spectra}); results for all sixteen are given in Appendix~\ref{app:results}. The database records are not redistributed with the code; their identifiers are listed in Appendix~\ref{app:results} and the records can be obtained from the LSPDD database~\cite{LSPDD}.

\emph{Measurement range.} The featured spectra extend from about 308--327~nm to 859--900~nm (Table~\ref{tab:featured}). Emission outside the measured range is unavailable and is treated as zero (Section~\ref{sec:protocol}), so the reported $\eta_{\text{wall}}$ and $\eta_{\text{DRE}}$ are lower bounds, and $\eta_{\text{spec}}$ refers to the measured part of the spectrum. The size of the omitted emission depends on the technology: it is expected to be small for PC-white LED and fluorescent sources and large for thermal sources and for sources with spectra reaching into the infrared.

For thermal sources, results obtained from such a truncated range can be strongly misleading, as the modeled source described below shows (Fig.~\ref{fig:spectrum_halogen}). Evaluated over all of $\Omega$, it gives $\eta_{\text{wall}}=0.900$ (assumed), $\eta_{\text{spec}}=0.573$ and $\eta_{\text{DRE}}=0.515$. Evaluated only within 310--900~nm, as if it had been measured like the lamps above, it gives $\eta_{\text{wall}}=0.232$, $\eta_{\text{spec}}=0.561$ and $\eta_{\text{DRE}}=0.130$, a value within the range of the measured lamps (0.045--0.162) and about four times too low. Neither the luminous efficacy, which is unchanged at 25.6~lm/W because $V(\lambda)$ lies entirely within the window, nor $\eta_{\text{spec}}$ reveals the problem; the latter changes by only 2\,\% here because two effects nearly cancel (Appendix~\ref{app:truncation}), which is a property of this source and window and not a general one. Such a result is a valid lower bound but, without independent knowledge of the omitted emission, it is not an estimate of the daylight-referenced efficiency of the source, and it should not be compared with the values of sources measured over their full emission range (see also\ Section~\ref{sec:protocol}).

\emph{Modeled source.} As a full-range emitter, a Planckian radiator at 3200~K was evaluated over all of $\Omega$ with an assumed $P_{\text{optical}}/P_{\text{elec}}=0.9$ ($P_{\text{elec}}=100$~W), that is, $\eta_{\text{wall}}=0.9$ by assumption. This is a modeled source, not a measurement: real tungsten filaments are greybodies, filament temperatures of halogen products span roughly 2700--3200~K, and infrared-reflecting coatings may also alter the radiant balance while the infrared range of emission extends beyond 4000~nm. The Planckian model and assumed wall-plug efficiency, nevertheless, provide a physically realistic approximation to the behavior of a tungsten filament. Its values are here to illustrate how the metric operates and are not claims about real halogen lamps.

\subsection{Results}
\begin{figure*}[t]
\includegraphics[width=\linewidth]{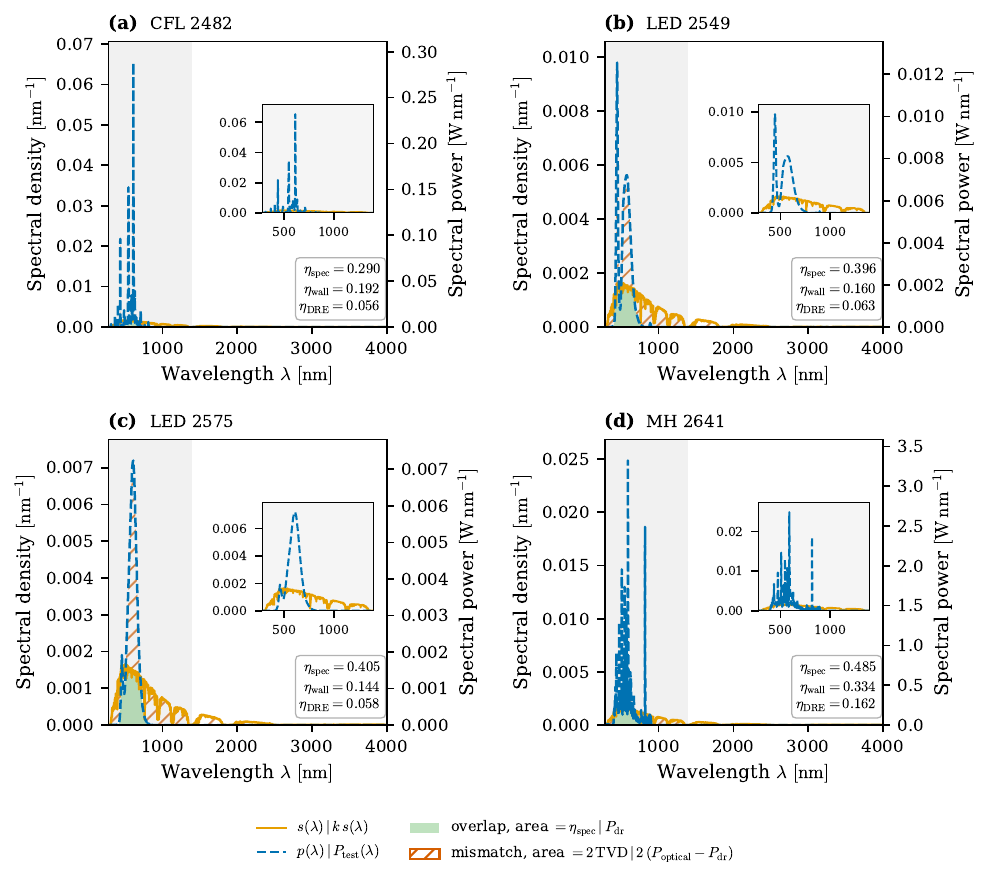}
\caption{Spectral fit against AM1.5G for the featured lamps: (a) CFL 2482, (b) LED 2549, (c) LED 2575, (d) MH 2641. Dashed blue: normalized lamp spectrum $p(\lambda)$ (left axis) and absolute spectrum $P_{\text{test}}(\lambda)$ (right axis). Solid amber: AM1.5G density $s(\lambda)$ and the same reference scaled to the lamp's total power, $k\,s(\lambda)$ with $k=P_{\text{optical}}$. Green: overlap $\min(p,s)$, whose area is $\eta_{\text{spec}}$ (left) and $P_{\text{dr}}$ (right). Hatched: mismatch $|p-s|$, whose area is $2\,\mathrm{TVD}$ (left axis) and $2\,(P_\mathrm{optical}-P_\mathrm{dr})$ (right axis), i.e.\ twice the power not counted toward $P_\mathrm{dr}$, since $|p-s|$ integrates both the excess and the equal deficit relative to the scaled reference. All axes are linear; vertical ranges differ between panels. Gray background shading and inset figures highlight the range of 280--1400~nm.}
\label{fig:spectra}
\end{figure*}

\begin{figure*}[t]
\includegraphics[width=\linewidth]{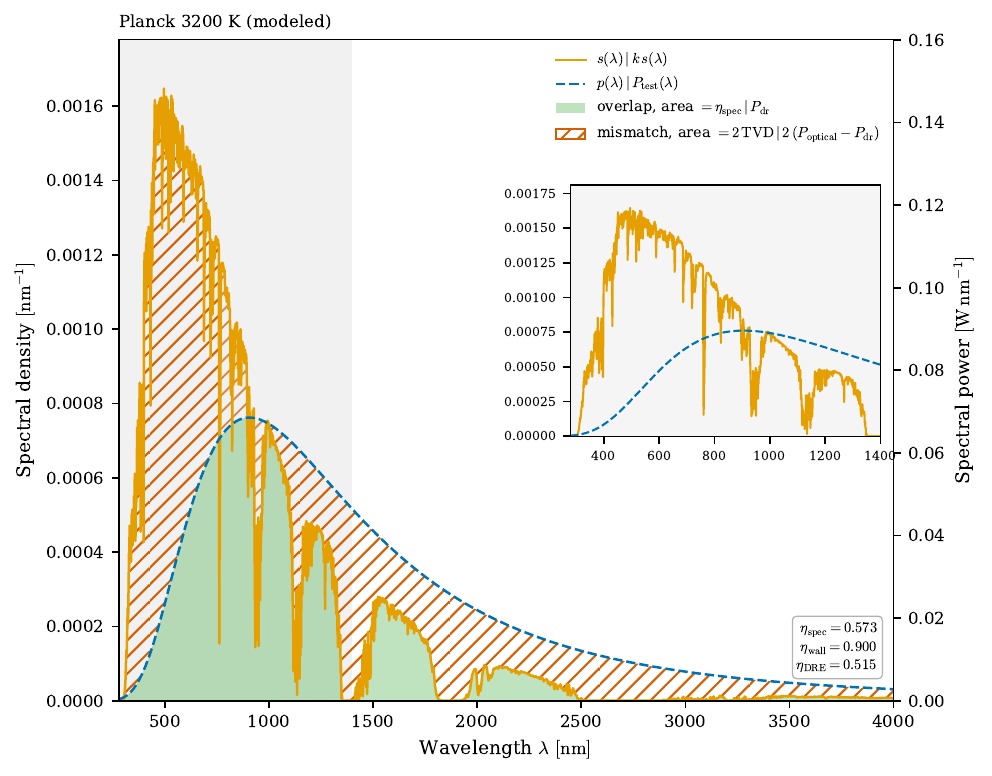}
\caption{Spectral fit against AM1.5G for the modeled 3200 K blackbody source. Dashed blue: normalized lamp spectrum $p(\lambda)$ (left axis) and absolute spectrum $P_{\text{test}}(\lambda)$ (right axis). Solid amber: AM1.5G density $s(\lambda)$ and the same reference scaled to the lamp's total power, $k\,s(\lambda)$ with $k=P_{\text{optical}}$. Green: overlap $\min(p,s)$, whose area is $\eta_{\text{spec}}$ (left) and $P_{\text{dr}}$ (right). Hatched: mismatch $|p-s|$, whose area is $2\,\mathrm{TVD}$ (left axis) and $2\,(P_\mathrm{optical}-P_\mathrm{dr})$ (right axis), i.e.\ twice the power not counted toward $P_\mathrm{dr}$, since $|p-s|$ integrates both the excess and the equal deficit relative to the scaled reference. All axes are linear; vertical ranges differ between panels. Gray background shading and inset figures highlight the range of 280--1400~nm.}
\label{fig:spectrum_halogen}
\end{figure*}

Table~\ref{tab:featured} lists the results for the featured lamps and the modeled source, Fig.~\ref{fig:spectra} shows the spectral fit of the four measured lamps, and Fig.~\ref{fig:spectrum_halogen} shows the spectral fit of the modeled blackbody source. All four measured sources emit in the visible range but distribute it differently: the fluorescent lamp (a) and the metal-halide luminaire (d) are dominated by narrow lines that exceed the smooth reference by more than an order of magnitude over narrow windows, whereas the LED spectra (b, c) consist of two broader bands. The modeled blackbody source lacks power relative to the reference through much of the visible range while exceeding it substantially in the near-infrared.

\begin{table*}[t]
\caption{Results for the featured lamps (measured LSPDD records, anchored to the reported $K$ and $P_{\text{elec}}$) and for the modeled source. $K$ is the reported luminous efficacy (computed for the modeled source). ``Range'' is the measured wavelength range; the modeled source covers all of $\Omega$. For the measured lamps, $\eta_{\text{wall}}$ and $\eta_{\text{DRE}}$ are lower bounds (Section~\ref{sec:protocol}); for the modeled source, $\eta_{\text{wall}}$ is assumed.}
\label{tab:featured}
\begin{ruledtabular}
\begin{tabular}{llrrrcccc}
Record & Source & CCT (K) & $P_{\text{elec}}$ (W) & $K$ (lm/W) & Range (nm) & $\eta_{\text{wall}}$ & $\eta_{\text{spec}}$ & $\eta_{\text{DRE}}$ \\
\hline
2482 & CFL, Globe Twister & 2723 & 23 & 69.6 & 310.5--859.0 & 0.1916 & 0.2899 & 0.0556 \\
2549 & LED PAR20, LightingScience & 5155 & 8 & 52.5 & 327.0--899.5 & 0.1603 & 0.3960 & 0.0635 \\
2575 & LED PAR20, EnergySmart & 2524 & 7 & 47.1 & 315.0--866.0 & 0.1438 & 0.4054 & 0.0583 \\
2641 & MH high bay, Philips & 4259 & 400 & 100.0 & 308.0--899.5 & 0.3341 & 0.4845 & 0.1619 \\
Model & Planck 3200~K (modeled) & 3200 & 100 & 25.6 & 280--4000 & 0.9000 & 0.5727 & 0.5154 \\
\end{tabular}
\end{ruledtabular}
\end{table*}

Across the sixteen lamps, $\eta_{\text{spec}}$ ranges from 0.29 (CFL 2482) to 0.54 (halogen 2498), and the ten LED lamps lie between 0.38 and 0.45. These values are to be read against a ceiling set by the measured range $B$: since the evaluated spectra vanish outside $B$ and $\min(p,s)\le s$, Eq.~(\ref{eq:minenv}) gives $\eta_{\text{spec}}\le\int_B s\,d\lambda$. For the featured ranges of 308--900~nm this bound is 0.65--0.68, not unity. The wall-plug efficiency ranges from 0.097 to 0.334, and $\eta_{\text{DRE}}$ from 0.045 to 0.162. For anchored spectra $\eta_{\text{wall}}=K/\text{LER}$ (Appendix~\ref{app:anchoring}), where LER denotes luminous efficacy of radiation: the LED 2604 and the MH luminaire 2641 both report $K=100$~lm/W, but their luminous efficacies of radiation of about 330 and 299~lm/W$_{\text{opt}}$ give $\eta_{\text{wall}}=0.303$ and $0.334$.

The incandescent and halogen records have the lowest luminous efficacies (7.6--16~lm/W) but $\eta_{\text{DRE}}$ of 0.045--0.082, of the same order as LED lamps with $K=47$--50~lm/W (0.058--0.065). These records are the most strongly affected by the limited measurement range: for a Planckian radiator at 2400--3200~K, only about 10--26\,\% of the power within $\Omega$ lies between 310 and 900~nm (model calculation). Since the measured range of these records ends near 900~nm, their $\eta_{\text{wall}}$ and $\eta_{\text{DRE}}$ underestimate the true values by a large factor (see also Section~\ref{sec:sources} and Appendix~\ref{app:truncation}).

The modeled 3200~K radiator has $K=25.6$~lm/W, lower than every measured LED, CFL and MH lamp but higher than the incandescent and halogen records. Its $\eta_{\text{spec}}=0.573$ exceeds that of every measured lamp, and is not limited by the visible-band ceiling above because the model covers all of $\Omega$. Its $\eta_{\text{DRE}}=0.515$ is $0.9\,\eta_{\text{spec}}$ by construction, since $\eta_{\text{wall}}$ is assumed.

The coverage warning of Section~\ref{sec:protocol} was issued for one record (2645, Appendix~\ref{app:results}). For the three aforementioned thermal sources, the final reported values in the LSPDD data fall below $2\,\%$ for unknown reasons, thereby failing to trigger a coverage warning despite the strong likelihood that these sources do emit beyond 900~nm. For all sixteen records, the luminous efficacy recomputed from the resampled spectrum agrees with the reported value to within $0.07\,\%$ (Appendix~\ref{app:verification}, item 11). The results are conditional on the reported $K$ and $P_{\text{elec}}$ of each record and carry no uncertainty estimate.


\section{Conclusion}
\label{sec:conclusion}

This work introduced a radiometric framework to evaluate the electrical efficiency with which an electric light source produces optical radiation spectrally overlapping a standardized terrestrial solar reference across the optical domain $\Omega = [280\text{ nm}, 4000\text{ nm}]$.

Starting from a source's total unweighted radiant optical power ($P_{\text{optical}}$), the framework determines what fraction of that optical output can be regarded as spectrally coincident with daylight ($\eta_{\text{spec}}$) by evaluating the integrated common spectral component between the normalized source and reference density profiles.

Multiplying total radiant power by this overlap fraction isolates the daylight-conforming optical output power ($P_{\text{dr}}$).

Ultimately, Daylight-Referenced Efficiency ($\eta_{\text{DRE}} = P_{\text{dr}}/P_{\text{elec}} = \eta_{\text{wall}} \times \eta_{\text{spec}}$) measures the fraction of active electrical input power converted into spectrally reference-conforming radiant power, establishing a neutral, energy-based benchmark unconstrained by visual domain sensitivity functions.

The framework factorizes performance into primary physical quantities, each serving distinct engineering and specification roles:

\begin{enumerate}
    \item \textbf{Wall-Plug Efficiency (\texorpdfstring{$\eta_{\text{wall}}$}{eta\_wall}) [$\text{W}_{\text{optical}}/\text{W}_{\text{elec}}$]:} Quantifies the physical conversion of active electrical input power ($P_{\text{elec}}$) into unweighted radiant optical power ($P_{\text{optical}}$) restricted to domain $\Omega = [280\text{ nm}, 4000\text{ nm}]$. Adopting conventional optoelectronic definitions of radiant wall-plug efficiency \cite{Schubert2006} bounded to the solar reference domain, it establishes a hardware performance baseline without visual domain sensitivity penalties.
    \item \textbf{Daylight Spectral Fidelity (\texorpdfstring{$\eta_{\text{spec}}$}{eta\_spec}) [unitless, $0\text{--}1$]:} Measures the normalized geometric spectral match to the solar reference profile via TVD ($1 - \text{TVD}$), or equivalently, minimum-envelope overlap (Appendix~\ref{app:minenv}). Invariant to absolute power magnitude, it serves as an isolated benchmark for spectral shape fitting and determining a source's similarity to the daylight reference.
    \item \textbf{Daylight-Referenced Output Power (\texorpdfstring{$P_{\text{dr}}$}{P\_dr}) [$\text{W}$]:} Expresses the absolute radiant optical power emitted within the solar reference profile envelope. Because $P_{\text{dr}}$ is a physical power quantity independent of electrical drive power, it provides lighting specifiers, architectural and lighting designers, and photobiological engineers with a direct metric to specify and verify the required absolute flux of daylight-referenced radiation within an illuminated space.
    \item \textbf{Daylight-Referenced Efficiency (\texorpdfstring{$\eta_{\text{DRE}}$}{eta\_DRE}) [$\text{W}/\text{W}_{\text{elec}}$]:} Defines the net rate of daylight-equivalent radiant power generated per unit of active electrical power consumed ($\eta_{\text{DRE}} = \eta_{\text{wall}} \times \eta_{\text{spec}}$). It acts as the primary summary metric for comparative energy rating and standardization of broad-spectrum light sources.
\end{enumerate}

Key mathematical properties of the framework include reference-spectrum independence (allowing adaptation to AM1.5D, future population-weighted global daylight models, or other terrestrial solar profiles), domain-boundary robustness, and provable lower-bound monotonicity under spectral truncation ($\eta_{\text{wall}}^M \le \eta_{\text{wall}}$ and $\eta_{\text{DRE}}^M \le \eta_{\text{DRE}}$). The DRE framework is also domain-extensible. Open-source reference software has been provided to facilitate deployment.


\clearpage
\onecolumngrid
\section*{Declarations}
The author is a cofounder of Pixun Technologies Co. (Hermosa Beach, CA, USA), which develops broad-spectrum display and lighting products. However, this research was conducted entirely in a personal capacity on personal time, using personal resources, and without the use of company facilities, funding, or proprietary data.

\twocolumngrid
\bibliography{Veto_DRE}

\clearpage


\appendix

\section{Min-Envelope Form of the Spectral Fidelity}
\label{app:minenv}

\textbf{Identity.} Let $p,s \ge 0$ with $\int_\Omega p\,d\lambda = \int_\Omega s\,d\lambda = 1$. Then
\begin{equation}
\eta_{\text{spec}} = 1-\text{TVD}(p,s) = \int_\Omega \min\!\big(p(\lambda),s(\lambda)\big)\,d\lambda .
\label{eq:minenv}
\end{equation}

\emph{Proof.} For $a,b\ge 0$, $|a-b| = a+b-2\min(a,b)$. Integrating, $\int|p-s| = 1+1-2\int\min(p,s)$, and halving gives $\text{TVD} = 1-\int\min(p,s)$. \hfill$\square$

Multiplying Eq.~(\ref{eq:minenv}) by $P_{\text{optical}}$ and using $P_{\text{optical}}\,p = P_{\text{test}}$,
\begin{equation}
P_{\text{dr}} = \int_\Omega \min\!\big(P_{\text{test}}(\lambda),\, P_{\text{optical}}\,s(\lambda)\big)\,d\lambda .
\label{eq:wdr_minenv}
\end{equation}
Here $P_{\text{optical}}\,s(\lambda)$ is the solar reference rescaled to the source's own total power, and $P_{\text{dr}}$ is the part of the source's power lying beneath that envelope.

\textbf{Discrete form.} Let $Q[f]=\sum_i w_i f(\lambda_i)$ be any linear quadrature rule. Because the pointwise identity above is linear in $|p-s|$, $p$, $s$ and $\min(p,s)$, $Q[|p-s|]=Q[p]+Q[s]-2\,Q[\min(p,s)]$. The implementation normalizes with the same rule used for the TVD integral, so $Q[p]=Q[s]=1$ holds exactly and Eq.~(\ref{eq:minenv}) holds exactly in the discrete computation, not only in the continuum limit. For the modeled source of Appendix~\ref{app:numerics} on the 1~nm grid, the two sides agree to $10^{-16}$. If all $w_i\ge 0$ (true for Simpson's rule on a uniform grid), then $0\le Q[\min(p,s)]\le Q[p]=1$, so $0\le\eta_{\text{spec}}\le 1$ also holds numerically.

\section{Numerical Procedure}
\label{app:numerics}

\subsection{Reference data}
\emph{AM1.5G.} The global-tilt column of ASTM~G173-03 \cite{ASTMG173} was obtained from the copy bundled with \texttt{pvlib}~0.16.1 (the official NREL table, unmodified), rather than from a web download, so that the file is reproducible from a pinned package version (MD5 \texttt{b31870bd3500b58414100af7e4b0e04c}). The native grid is non-uniform (2002 wavelengths; 0.5~nm below 400~nm, 1~nm from 400 to 1700~nm, up to 5~nm above). Trapezoidal integration of the native table gives $1000.4\ \text{W/m}^2$; after PCHIP resampling and Simpson integration on the 1~nm grid it gives $1000.8\ \text{W/m}^2$ ($+0.04\,\%$). Since $s(\lambda)$ is normalized, only the spectral shape enters the metric.

\emph{$V(\lambda)$.} The CIE~1924 photopic function (2$^\circ$ observer, 1~nm steps, 360--830~nm) \cite{CIE018_2019} was exported from \texttt{colour-science}~0.4.7 (MD5 \texttt{7f741ff3b0f570bc7d8273f4d945ce15}) and checked against the CIE open-access dataset. It is taken as zero outside 360--830~nm. The SI defining constant is \(K_{\mathrm{cd}} = 683\ \mathrm{lm/W}\) at \(\nu = 540 \times 10^{12}\ \mathrm{Hz}\). For conventional photopic calculations, we use \(K_m = 683\ \mathrm{lm/W}\), the rounded value of the maximum spectral luminous efficacy at 555 nm.

\subsection{Grid and resampling}
The master grid is $\lambda_i = 280,281,\ldots,4000$~nm. Any input spectrum is processed in this order: (i) sort by wavelength and average duplicate wavelengths; (ii) PCHIP interpolation \cite{Fritsch1980} (no extrapolation) at the grid points; (iii) set grid points outside the native range to zero; (iv) set any remaining negative values to zero. PCHIP is monotone on every interval where the data are monotone, so for non-negative data the interpolant lies between neighboring data values and cannot become negative or overshoot a sharp peak. A cubic spline offers no such guarantee. Reference spectra, user files and anchored database spectra all pass through this single code path:
\begin{lstlisting}
interpolator = PchipInterpolator(native_wavelengths,
                                 native_values, extrapolate=False)
resampled = interpolator(master_grid)
resampled[np.isnan(resampled)] = 0.0  # unmeasured = absent
resampled = validation.clip_negative(resampled, name)
\end{lstlisting}
Resampling is point sampling: no instrument bandpass is deconvolved and no band-limiting filter is applied before decimation. For line-rich spectra measured at resolution finer than 1~nm, a finer uniform grid can be passed to the code (the integrals are grid-independent in form). The calculation requires only a uniform grid and does not assume 1~nm or the full $\Omega$.

\subsection{Quadrature}
Every integral is evaluated with the composite Simpson rule (\texttt{scipy.integrate.simpson}) on the uniform grid. The default grid has an odd number of points (3721), so the rule applies without end corrections, and its weights $\tfrac{\Delta\lambda}{3}(1,4,2,\ldots,4,1)$ are all positive, a property used in Appendices~\ref{app:minenv} and \ref{app:truncation}. The core of the calculation is:
\begin{lstlisting}
p_optical = simpson(p_test_raw, x=wavelengths)
eta_wall  = p_optical / p_elec
s_integral = simpson(s_am15g, x=wavelengths)
p_lambda  = p_test_raw / p_optical
s_lambda  = s_am15g / s_integral
tvd       = 0.5 * simpson(np.abs(p_lambda - s_lambda),
                          x=wavelengths)
eta_spec  = 1.0 - tvd
p_dr      = p_optical * eta_spec
eta_dre   = eta_wall * eta_spec
\end{lstlisting}
The integrand $|p-s|$ is only piecewise smooth (it has kinks where the two densities cross), so Simpson's rule is not high-order for that term. This is immaterial in practice: replacing Simpson's rule by the trapezoidal rule changes $\eta_{\text{DRE}}$ by $6.8\times10^{-4}$ and $8.7\times10^{-4}$ (relative) for the two sources below. For a smooth integrand, Simpson's rule on the 1~nm grid reproduces an adaptive quadrature of the Planck function ($T=5778$~K, 280--4000~nm) to a relative difference of $6\times10^{-12}$ (trapezoidal: $6\times10^{-7}$).

\subsection{Preprocessing and reporting}
Negative values are replaced by zero (with a warning stating the count and the most negative value) at three points: on the inputs of every calculation, after interpolation, and on relative spectra before anchoring. If a file's measured range extends beyond $\Omega$, the fraction of its own measured power lying outside $\Omega$ is computed exactly from the data and reported; this power is excluded from all quantities. Leading header lines and the column delimiter of instrument files are detected automatically; the first two numeric columns are read as (wavelength~[nm], value).

Note that hard-clipping negative spectral values can rectify zero-mean detector noise into a positive bias, particularly across broad wavelength regions with negligible true emission, potentially inflating integrated optical power and distorting normalized spectral overlap metrics.

\subsection{Grid-resolution dependence}
Table~\ref{tab:grid} shows $\eta_{\text{DRE}}$ computed at several grid spacings for (a) a \emph{modeled} 3200~K blackbody with assumed $P_{\text{optical}}=90$~W and $P_{\text{elec}}=100$~W, and (b) a \emph{measured} LED (LSPDD record 2575, EnergySmart PAR20, $P_{\text{elec}}=7$~W, $47.1$~lm/W; \cite{LSPDD}). Deviations are relative to the 0.25~nm result. At the default 1~nm spacing they are $4\times10^{-4}$ and $9\times10^{-4}$; at 10~nm they are of order $10^{-3}$--$2\times10^{-3}$, and at 20~nm up to $10^{-2}$. The convergence is not monotone. Since Simpson-versus-trapezoid differences are of similar size, we attribute the residual variation mainly to resampling of structured features (the atmospheric absorption bands of AM1.5G, narrow LED features) rather than to quadrature error.

\begin{table}[b]
\caption{Dependence of $\eta_{\text{DRE}}$ on grid spacing $\Delta\lambda$. Deviation is relative to $\Delta\lambda=0.25$~nm. The blackbody is modeled, the LED is measured (see text).}
\label{tab:grid}
\begin{ruledtabular}
\begin{tabular}{lrrrr}
$\Delta\lambda$ (nm) & Blackbody & Dev. & LED & Dev. \\
\hline
0.25 & 0.5152  & ---  & 0.05835 & ---  \\
0.5  & 0.5153  & $1.2\times10^{-4}$ & 0.05834 & $3.5\times10^{-5}$ \\
1    & 0.5154  & $4.2\times10^{-4}$ & 0.05830 & $8.5\times10^{-4}$ \\
2    & 0.5141  & $2.1\times10^{-3}$ & 0.05853 & $3.2\times10^{-3}$ \\
5    & 0.5145  & $1.3\times10^{-3}$ & 0.05838 & $5.5\times10^{-4}$ \\
10   & 0.5141  & $2.2\times10^{-3}$ & 0.05843 & $1.4\times10^{-3}$ \\
20   & 0.5170  & $3.5\times10^{-3}$ & 0.05895 & $1.0\times10^{-2}$ \\
\end{tabular}
\end{ruledtabular}
\end{table}

\section{Spectral Truncation}
\label{app:truncation}

\textbf{Setting.} Let $P(\lambda)\ge 0$ be the true emission of a source on $\Omega$ and $M\subseteq\Omega$ the set of measured wavelengths. Under the zero-fill convention the evaluated spectrum is $P^M(\lambda)=P(\lambda)\,\mathbf 1_M(\lambda)$; $P_{\text{elec}}$ and the reference $s(\lambda)$ are unchanged. Emission outside $\Omega$ is not part of the metric by definition, so a measurement with $M\supseteq\Omega$ is complete.

\textbf{Proposition.} $\eta_{\text{wall}}^M\le\eta_{\text{wall}}$ and $\eta_{\text{DRE}}^M\le\eta_{\text{DRE}}$.

\emph{Proof.} Since $0\le P^M\le P$ pointwise, $P_{\text{optical}}^M=\int P^M\le\int P=P_{\text{optical}}$, which gives the statement for $\eta_{\text{wall}}$ after division by $P_{\text{elec}}$. By Eq.~(\ref{eq:wdr_minenv}),
\begin{equation}
P_{\text{dr}}^M=\int_\Omega \min\!\big(P^M,\,P^M_{\text{optical}}\,s\big)\,d\lambda .
\end{equation}
The function $(x,y)\mapsto\min(x,y)$ is non-decreasing in each argument, and $s\ge 0$. Hence $P^M\le P$ and $P^M_{\text{optical}}s\le P_{\text{optical}}s$ imply, pointwise, $\min(P^M,P^M_{\text{optical}}s)\le\min(P,P_{\text{optical}}s)$. Integrating yields $P^M_{\text{dr}}\le P_{\text{dr}}$, and dividing by $P_{\text{elec}}$ gives the claim for $\eta_{\text{DRE}}$.

The same argument holds for the discrete computation, because the quadrature weights are positive (Appendix~\ref{app:numerics}) and Eq.~(\ref{eq:wdr_minenv}) is exact for the discrete rule (Appendix~\ref{app:minenv}). The result is independent of the source's spectral shape.

\textbf{Spectral fidelity.} Since $\eta_{\text{spec}}=P_{\text{dr}}/P_{\text{optical}}$ is a ratio of two quantities that both decrease under truncation, it can move in either direction. Table~\ref{tab:trunc} shows this for the modeled 3200~K blackbody ($P_{\text{elec}}=100$~W) with the long-wavelength end cut at $\lambda_c$: $\eta_{\text{spec}}$ first rises (the removed tail was farther from the daylight shape than the retained part) and then falls once the cut reaches the visible and near-IR, where the source and the reference overlap. A randomized check supports the proposition and the two-sided behavior: for 300 spectra (random temperature, random multiplicative structure, random power, random measured window within $\Omega$), the inequalities for $\eta_{\text{wall}}$ and $\eta_{\text{DRE}}$ were never violated, while $\eta_{\text{spec}}$ increased in 33 cases and decreased in 267.

\textbf{Why $\eta_{\text{spec}}$ can be insensitive to truncation.} For a measured window $B$, the evaluated density is $p^B=p\,\mathbf 1_B/m_p$ with $m_p=\int_B p\,d\lambda$, and $\eta_{\text{spec}}^B=\int_B\min(p^B,s)\,d\lambda\le\int_B s\,d\lambda$. Truncation therefore has two opposing effects: overlap outside $B$ is lost, and the retained shape is renormalized to unit area, which raises its density by $1/m_p$ and increases the overlap inside $B$. For the modeled 3200~K source and $B=310$--$900$~nm, $m_p=0.258$; the overlap outside $B$ (0.317, equal to the reference's entire power fraction beyond 900~nm, since the source lies above the reference throughout the near-infrared) is lost, while the overlap inside $B$ rises from 0.256 to 0.561 (82\,\% of the ceiling 0.683), so that $\eta_{\text{spec}}$ changes only from 0.573 to 0.561. The near-cancellation is specific to this source and window: for the same source, $\eta_{\text{spec}}=0.490$ for $B=380$--$780$~nm, $0.422$ for $400$--$700$~nm, and $0.317$ for $900$--$4000$~nm.

\begin{table}[b]
\caption{Effect of cutting the long-wavelength end of a modeled 3200~K blackbody at $\lambda_c$ (all quantities recomputed on the truncated spectrum with fixed $P_{\text{elec}}$).}
\label{tab:trunc}
\begin{ruledtabular}
\begin{tabular}{lrrr}
$\lambda_c$ (nm) & $\eta_{\text{wall}}$ & $\eta_{\text{spec}}$ & $\eta_{\text{DRE}}$ \\
\hline
4000 (full) & 0.9000 & 0.5727 & 0.5154 \\
2500 & 0.8084 & 0.5933 & 0.4796 \\
2000 & 0.7263 & 0.5941 & 0.4315 \\
1100 & 0.3656 & 0.6009 & 0.2197 \\
900  & 0.2324 & 0.5611 & 0.1304 \\
780  & 0.1517 & 0.5017 & 0.0761 \\
\end{tabular}
\end{ruledtabular}
\end{table}

\textbf{Detection heuristic.} The measured spectrum's value at each range edge is compared with its own peak. If an edge lies strictly inside $\Omega$ and the value exceeds 2\,\% of the peak, a dedicated warning is issued stating the bounds above. If the edge coincides with or lies beyond a boundary of $\Omega$, nothing relevant to the metric is missing and only an informational note is printed:
\begin{lstlisting}
edge_low_ratio  = native_values[0]  / peak
edge_high_ratio = native_values[-1] / peak
# edge > threshold and inside Omega -> SpectralCoverageWarning
# edge > threshold and at/beyond Omega -> informational note only
\end{lstlisting}
The heuristic cannot detect a weak but broad tail that has fallen below 2\,\% of the peak yet still carries appreciable power. It does not replace reporting the measured range.

\textbf{Interaction with anchoring.} The proposition assumes that the absolute scale of the retained spectrum is that of the true source. For a relative spectrum anchored through $V(\lambda)$ (Appendix~\ref{app:anchoring}), this holds if the truncated region lies outside 360--830~nm. If the cut removes part of the visible band, the anchor is computed from the truncated flux and rescales the retained part upward, so the bounds are no longer guaranteed.

\section{Anchoring of Relative Spectra}
\label{app:anchoring}

Some public databases (e.g.\ \cite{LSPDD}) provide a relative spectral shape $p_{\text{rel}}(\lambda)$ of unknown scale together with a reported luminous efficacy $K$ [lm/W$_{\text{elec}}$] and electrical power $P_{\text{elec}}$. The reported luminous flux is $\Phi_v=K\,P_{\text{elec}}$. If $p_{\text{rel}}$ were in W/nm it would imply $\Phi_v^{\text{rel}}=K_m\int p_{\text{rel}}V\,d\lambda$. The absolute spectrum is therefore
\begin{equation}
P_{\text{test}}(\lambda)=c\,p_{\text{rel}}(\lambda),\qquad c=\frac{K\,P_{\text{elec}}}{K_m\int p_{\text{rel}}(\lambda)V(\lambda)\,d\lambda}.
\end{equation}
\textbf{Invariance.} Replacing $p_{\text{rel}}$ by $\alpha\,p_{\text{rel}}$ (peak-normalized, area-normalized or any other convention) replaces $c$ by $c/\alpha$, so $P_{\text{test}}$ is unchanged. The anchoring is therefore independent of the database's normalization convention.

\textbf{Consequences.} With the luminous efficacy of radiation $\text{LER}=K_m\int p_{\text{rel}}V\,d\lambda\big/\!\int p_{\text{rel}}\,d\lambda$,
\begin{equation}
\eta_{\text{wall}}=\frac{c\int p_{\text{rel}}\,d\lambda}{P_{\text{elec}}}=\frac{K}{\text{LER}} ,
\end{equation}
so for anchored spectra $P_{\text{elec}}$ cancels, and $\eta_{\text{wall}}$ and $\eta_{\text{DRE}}$ depend only on the reported $K$ and the relative shape. A relative uncertainty in $K$ propagates one-to-one into $\eta_{\text{wall}}$ and $\eta_{\text{DRE}}$ and does not affect $\eta_{\text{spec}}$. For the LED of Appendix~\ref{app:numerics}, $\text{LER}=327.6$~lm/W$_{\text{opt}}$ and $K/\text{LER}=0.1438$, matching the computed $\eta_{\text{wall}}$; recomputing with $P_{\text{elec}}$ set to 3.5, 7 and 14~W gives identical $\eta_{\text{wall}}$ and $\eta_{\text{DRE}}$.

\textbf{Implementation.} Anchoring is performed on the native wavelength grid, with $V(\lambda)$ evaluated at the same points, and the resulting absolute spectrum is then passed through the standard resampling of Appendix~\ref{app:numerics}:
\begin{lstlisting}
implied_relative_lumens = K_m * simpson(
    native_relative_values * v_lambda_native,
    x=native_wavelengths)
reported_lumens = luminous_efficacy_lm_per_w * p_elec_w
scale_factor = reported_lumens / implied_relative_lumens
native_absolute_values = scale_factor * native_relative_values
\end{lstlisting}
Anchoring fails with an error if the relative spectrum has no overlap with 360--830~nm.

\textbf{Assumptions.} The result is valid to the extent that (i) the reported $K$ and $P_{\text{elec}}$ refer to the same operating point as the spectrum; (ii) the relative spectrum is representative of the total emitted flux (no strong angular dependence of color); (iii) $K$ was computed with the CIE~1924 $V(\lambda)$ and the same $K_m$; (iv) the measured range covers the visible band; and (v) emission outside the measured range is zero (otherwise $\eta_{\text{wall}}$ and $\eta_{\text{DRE}}$ are lower bounds, subject to the caveat in Appendix~\ref{app:truncation}). Records lacking either $K$ or $P_{\text{elec}}$ cannot be anchored and are excluded.

\section{Verification}
\label{app:verification}

Table~\ref{tab:verif} lists the checks. Tolerances shown are those of the automated test suite distributed with the code; where an observed value is given, it was obtained by running the same computation.

\begin{table*}[t]
\caption{Verification checks. ``Tol.'' is the tolerance used in the automated test.}
\label{tab:verif}
\small
\setlength{\tabcolsep}{4pt}
\begin{ruledtabular}
\begin{tabular}{c c c c}
\# & \parbox[t]{0.28\textwidth}{\raggedright Check} & \parbox[t]{0.28\textwidth}{\raggedright Expected} & \parbox[t]{0.28\textwidth}{\raggedright Observed / tol.} \\
\hline
1 & \parbox[t]{0.28\textwidth}{\raggedright Source $\propto$ reference (any scale)} & \parbox[t]{0.28\textwidth}{\raggedright $\eta_{\text{spec}}=1$, $\eta_{\text{DRE}}=\eta_{\text{wall}}$} & \parbox[t]{0.28\textwidth}{\raggedright tol.\ $10^{-9}$} \\
2 & \parbox[t]{0.28\textwidth}{\raggedright Gaussian lines ($\sigma~= 10\,\mathrm{nm}$) at 350~nm (stand-in reference) and 950~nm (test)} & \parbox[t]{0.28\textwidth}{\raggedright $\eta_{\text{spec}}=10^{-16}$} & \parbox[t]{0.28\textwidth}{\raggedright tol.\ $10^{-3}$} \\
3 & \parbox[t]{0.28\textwidth}{\raggedright Scaling $P_{\text{test}}\to 500\,P_{\text{test}}$} & \parbox[t]{0.28\textwidth}{\raggedright $\eta_{\text{spec}}$ unchanged} & \parbox[t]{0.28\textwidth}{\raggedright tol.\ $10^{-9}$} \\
4 & \parbox[t]{0.28\textwidth}{\raggedright Normalization of $p$, $s$} & \parbox[t]{0.28\textwidth}{\raggedright $\int p=\int s=1$} & \parbox[t]{0.28\textwidth}{\raggedright tol.\ $10^{-6}$} \\
5 & \parbox[t]{0.28\textwidth}{\raggedright $\eta_{\text{DRE}}$ vs.\ $P_{\text{dr}}/P_{\text{elec}}$} & \parbox[t]{0.28\textwidth}{\raggedright equal} & \parbox[t]{0.28\textwidth}{\raggedright tol.\ $10^{-9}$} \\
6 & \parbox[t]{0.28\textwidth}{\raggedright Simpson vs.\ adaptive quadrature, Planck 5778~K} & \parbox[t]{0.28\textwidth}{\raggedright agree} & \parbox[t]{0.28\textwidth}{\raggedright $6\times10^{-12}$ (tol.\ $10^{-3}$)} \\
7 & \parbox[t]{0.28\textwidth}{\raggedright 1~W line at 555~nm, photopic flux} & \parbox[t]{0.28\textwidth}{\raggedright $683$~lm ($K_m$)} & \parbox[t]{0.28\textwidth}{\raggedright tol.\ $10^{-3}$ rel.} \\
8 & \parbox[t]{0.28\textwidth}{\raggedright Bundled data} & \parbox[t]{0.28\textwidth}{\raggedright $V(555)=1$, support 360--830~nm; AM1.5G $\approx1000$~W/m$^2$} & \parbox[t]{0.28\textwidth}{\raggedright $1.0$; 360--830~nm; 1000.4~W/m$^2$ native (test range 800--1200~W/m$^2$)} \\
9 & \parbox[t]{0.28\textwidth}{\raggedright Discrete identity, Eq.~(\ref{eq:minenv})} & \parbox[t]{0.28\textwidth}{\raggedright $\eta_{\text{spec}}=Q[\min(p,s)]$} & \parbox[t]{0.28\textwidth}{\raggedright agree to $10^{-16}$ (tol.\ $10^{-12}$)} \\
10 & \parbox[t]{0.28\textwidth}{\raggedright Randomized truncation (300 spectra)} & \parbox[t]{0.28\textwidth}{\raggedright no violation of bounds for $\eta_{\text{wall}},\eta_{\text{DRE}}$} & \parbox[t]{0.28\textwidth}{\raggedright 0 violations} \\
11 & \parbox[t]{0.28\textwidth}{\raggedright LED round trip (anchor $\to$ resample $\to$ $K$)} & \parbox[t]{0.28\textwidth}{\raggedright $47.1$~lm/W} & \parbox[t]{0.28\textwidth}{\raggedright $47.101$, $+0.002\,\%$ (tol.\ $1\,\%$)} \\
\end{tabular}
\end{ruledtabular}
\end{table*}

The round trip (item 11) takes the relative LED spectrum, anchors it to the reported $K$ and $P_{\text{elec}}$ on the native grid, resamples it to the 1~nm master grid (including zero-fill) and recomputes $K$ from the resampled spectrum. Because the anchoring forces agreement on the native grid, the test verifies that resampling, zero-fill, clipping and Simpson integration on the master grid preserve the luminous flux; it does not validate the absolute radiometric scale of the underlying lamp, which rests on the reported values (Appendix~\ref{app:anchoring}). No independent absolutely calibrated broadband spectrum covering all of $\Omega$ was available for an end-to-end comparison. The remaining tests of the suite cover input handling (header and delimiter detection, duplicate wavelengths, zero-fill, PCHIP interpolation of a smooth function, negative-value clipping, validation errors) and the warning logic of Appendix~\ref{app:truncation}. The full suite (111 tests) passes in a clean virtual environment (103 passes and 8 skips without the tested LSPDD records \cite{LSPDD}, which are not contained in the reference implementation).

\section{Complete Results}
\label{app:results}

Table~\ref{tab:all} lists the results for all sixteen evaluated LSPDD records \cite{LSPDD} and the modeled source. Record numbers refer to the database entries; the spectra are not redistributed with the code, and each record is available from LSPDD under its own license (CC BY-NC-ND). $K_{\text{rep}}$ is the reported and $K_{\text{comp}}$ the recomputed luminous efficacy; $\Delta$ is their relative difference.

\begin{table*}[t]
\caption{Results for all evaluated records. (a)~Modeled, not measured. (b)~Assumed. (c)~Coverage warning issued (Appendix~\ref{app:truncation}). $\eta_{\text{wall}}$ and $\eta_{\text{DRE}}$ of the measured records are lower bounds (Section~\ref{sec:protocol}). Units: K for CCT, W for \(P_{\text{elec}}\), lm/W for $K$, \% for $\Delta$, and nm for Range.}
\label{tab:all}
\begin{ruledtabular}
\begin{tabular}{lllrrrrrrrrrr}
Record & Type & Source & CCT & $P_{\text{elec}}$ & $K_{\text{rep}}$ & $K_{\text{comp}}$ & $\Delta$ & Range & $\eta_{\text{wall}}$ & $\eta_{\text{spec}}$ & $\eta_{\text{DRE}}$ \\
\hline
2469 & LED & CREE BR30 & 2691 & 9.5 & 68.4 & 68.40 & +0.002 & 300--899 & 0.2021 & 0.4094 & 0.0827 \\
2482 & CFL & Globe Twister & 2723 & 23 & 69.6 & 69.59 & $-$0.020 & 310.5--859 & 0.1916 & 0.2899 & 0.0556 \\
2493 & Incand. & Sylvania A19 & 2400 & 25 & 7.6 & 7.60 & +0.007 & 307--899.5 & 0.0969 & 0.4650 & 0.0451 \\
2494 & Halogen & Globe Vanity & 2835 & 43 & 14.0 & 14.00 & $-$0.001 & 305--899.5 & 0.1285 & 0.5249 & 0.0674 \\
2498 & Halogen & Globe R30 & 3069 & 72 & 16.0 & 16.00 & +0.001 & 300--899.5 & 0.1524 & 0.5410 & 0.0824 \\
2504 & CFL & Globe Twister & 3692 & 13 & 69.2 & 69.15 & $-$0.070 & 304.5--899.5 & 0.2500 & 0.4479 & 0.1120 \\
2531 & LED & Standard PAR20 & 5015 & 8 & 61.0 & 61.00 & +0.004 & 380--889 & 0.1778 & 0.3842 & 0.0683 \\
2549 & LED & LightingSci. PAR20 & 5155 & 8 & 52.5 & 52.50 & +0.001 & 327--899.5 & 0.1603 & 0.3960 & 0.0635 \\
2562 & LED & Sylvania A19 & 4987 & 4 & 50.0 & 50.00 & +0.003 & 301--899.5 & 0.1535 & 0.4222 & 0.0648 \\
2575 & LED & EnergySmart PAR20 & 2524 & 7 & 47.1 & 47.10 & +0.002 & 315--866 & 0.1438 & 0.4054 & 0.0583 \\
2592 & LED & Globe A19 & 5060 & 4 & 50.0 & 50.00 & +0.004 & 300--899.5 & 0.1540 & 0.4222 & 0.0650 \\
2604 & LED & Tlsanli A19 & 5114 & 5 & 100.0 & 99.99 & $-$0.005 & 300--899.5 & 0.3031 & 0.4220 & 0.1279 \\
2641 & Metal h. & Philips High Bay & 4259 & 400 & 100.0 & 99.95 & $-$0.053 & 308--899.5 & 0.3341 & 0.4845 & 0.1619 \\
2645 & LED & RAB Ceiling Light\textsuperscript{c} & 5056 & 14 & 88.0 & 88.01 & +0.007 & 300--898 & 0.2496 & 0.3954 & 0.0987 \\
2650 & LED & RAB VEK-2 & 4888 & 52 & 72.0 & 72.00 & +0.001 & 305--898 & 0.2035 & 0.3879 & 0.0790 \\
2742 & LED & GE Refresh & 4759 & 8.5 & 94.0 & 93.99 & $-$0.006 & 303--899.5 & 0.3186 & 0.4460 & 0.1421 \\
Model & Blackb. & Planck 3200~K\textsuperscript{a} & 3200 & 100 & --- & 25.58 & --- & 280--4000 & 0.9000\textsuperscript{b} & 0.5727 & 0.5154 \\
\end{tabular}
\end{ruledtabular}
\end{table*}

\end{document}